%Paper: hep-ph/9311310
%From: ranfone@ifae.es
%Date: 18 Nov 1993 13:02:25 +0200

%Use just "plain Tex"
\magnification=1200
\font\text=cmr10
\font\it=cmti10

\font\title=cmbx10 scaled \magstep2

 \newcount\notenumber
\def\clearnotenumber{\notenumber=0}
\def\note{\advance\notenumber by 1
\footnote{$^{\the\notenumber}$}}
\clearnotenumber
\baselineskip=17pt
\text
{\baselineskip9pt\vbox{\hbox{UAB-FT-324}}}
\bigskip

\centerline{\title The Neutrino Mass Spectrum in the Supersymmetric}
\centerline{\title Flipped SU(5)$\bigotimes$U(1) GUT Model}

\vskip 1.5 cm

\centerline{\bf Stefano Ranfone}

\medskip

\centerline{\it Instituto de Fisica de Altas Energias, Grupo de Fisica
Teorica}
\centerline{\it Universitat Autonoma de Barcelona}
 \centerline{\it 08193 Bellaterra, Barcelona, Spain}

\vskip 4 cm

\centerline{\bf Abstract}

In the context of the supersymmetric flipped $SU(5)\otimes
U(1)$  GUT model, we have studied in detail the neutrino mass spectrum,
obtaining an approximate formula for the corresponding effective
three-generation mass matrix. It is found that at least two neutrinos are
expected to have extremely small masses of no particular interest. The third
neutrino, on the other hand, may eventually get a mass of order $10^{-3}$ eV,
of relevance for the solution of the solar neutrino problem with the MSW
mechanism, as a consequence of vacuum-expectation-values at the GUT mass scale
for some of the scalar partners of the right-handed neutrinos.

\vskip 2 cm

\centerline{November 1993}
\vfill\eject

In the last few years there has been a considerable activity in the area of the
GUT models derivable from superstring theories. Among the others,
the most celebrated is perhaps the so-called ``flipped" $SU(5)\otimes U(1)$
model, in its minimal supersymmetric N=1 GUT version [1]. It has several nice
features, which make it one of the most promising GUT models which have been
constructed so far. First of all, it does not require the presence of the
adjoint  Higgs representation (not allowed in models based on string theories
with Kac-Moody level K=1) for obtaining the correct spontaneous symmetry
breaking down to the standard model. Furthermore, it produces
automatically the so-called doublet-triplet mass splitting, essential for
avoiding an unplausible fast proton decay mediated by the exchange of light
color-triplet scalars. There are also extended  ``string"-type versions of
this model [1,2], which have been extensively studied in the literature.
However, most of them reduce at low-energy to the minimal version we are
considering in the present paper.  In view of its successes, it has become of
primary importance the study of its
 consequences and predictions, and in particular its expectations for the
physics of fermion masses. As
far as the charged sector is concerned, in the {\sl flipped} model the
quark and the charged-lepton masses turn out to be uncorrelated (unless one
embeds the model in $SO(10)$), so that one does not expect any constraint from
the observed mass spectra.  On the other hand, the model mantains the equality
(at the GUT mass scale, $M_G$) of the up-quark and the Dirac
neutrino mass matrices, implying therefore the need of a seesaw-type of
mechanism for the suppression of the neutrino masses [12]. The  study of the
possible seesaw scenarios which may be implemented in the model, both in the
supersymmetric and in the non-supersymmetric case, has
been given in refs.[3,4]. Also the present paper is devoted to the discussion
of the possible neutrino mass spectra which may arise in the flipped model, but
here we shall be able to carry out a more detailed study, at a three-generation
level, and we shall derive a general formula for the corresponding mass matrix,
holding also in presence of non-vanishing vacuum-expectation-values (VEVs) for
the scalar partners of the right-handed (RH) neutrinos.

The introduction of these VEVs, $<\nu^c>$,  was motivated in ref.[5], by the
study of the mass relations between the $d$-type quarks and the charged leptons
(not holding in the flipped case) which one obtains in the context
of the $SU(4)\otimes O(4)$ model\footnote{$^{1}$}{We recall that, as shown in
refs.[3,4],
the $SU(4)\otimes O(4)$ and the flipped $SU(5)\otimes U(1)$ models give in
general very similar results in the neutral lepton sector.}.  As it is well
known, these relations, also valid in the standard $SU(5)$ GUT model,
 are consistent with the actual masses only for the third generation,
corresponding to the famous relation (at a low-energy scale) $m_b \simeq 3
m_{\tau}$, where the factor of 3 arises from the different mass renormalization
in the two fermion sectors.  Leaving aside the problem with the masses of the
first generation, in view of possible large effects  of non-perturbative QCD on
the light $d$-quark mass, one may expect a better success for the second
generation, since  $m_s$ is larger (or of the same order) than
$\Lambda_{QCD}$. A way for preserving the successful relation between $m_b$ and
$m_{\tau}$, while improving the one involving $m_s$ and $m_{\mu}$, was
suggested
in ref.[5], by assuming non-vanishing VEVs for
the RH sneutrinos. In
particular, it was found that in order to fit $m_s$ and $m_{\mu}$ to their
actual values, $<\nu_{\mu}^c>$ had to be set at the GUT mass scale, $M_G$. The
general  consequences of this result on the structure of the neutrino
mass spectrum were already studied in ref.[4], where it was shown that in this
case sizeable ({\it i.e.,} non-negligibly small) masses could be obtained
without introducing non-renormalizable terms.

Before proceeding to the discussion of the neutrino sector, let us briefly
review the main ingredients of the flipped
$SU(5)\otimes U(1)$ supersymmetric GUT model. Its superfield content
 has been given in the Table of ref.[3], to which we also refer for our
notations.  The matter superfields, which accomodate for each generation the
fifteen fermions of the Standard Model (SM) plus the right-handed neutrino
$\nu^c$, form a 16-dimensional spinorial representation of $SO(10)$,
$\,\,\bf F(10,1)\,\oplus {\bar f}({\bar 5},-3)\,\oplus l^c(1,5)$. The particles
fit in these supermultiplets as in the Standard $SU(5)$ model (to which is
added the RH neutrino), but with the exchange $u^{(c)}\leftrightarrow d^{(c)}$,
and ${\nu}^{(c)}\leftrightarrow e^{(c)}$, which is the reason for the term
``flipped".  Therefore, we have ${\bf F}\sim(u,d,d^c,\nu^c)$, $\,{\bf
f}\sim(u^c,\nu,e)\,$ and $\,{\bf l^c}\sim e^c$. The gauge group of the model
is spontaneously broken down to the SM at the GUT mass scale, $M_G = 10^{16}$
GeV, by the VEV of the neutral components
of two Higgs superfields $\bf H(10,1)$ and $\bf{\bar{H}}({\bar{
10}},-1)$, $\,\,\,<H>\equiv <\nu^c_H>\,\,$ and $\,\,<{\bar{H}}>\equiv
<{\bar\nu}^c_H>$. From the minimization of the F and the D terms one finds that
these two VEVs must be equal, so that we may set
$<\nu^c_H>=<{\bar\nu}^c_H>\equiv M_G$. Through this spontaneous symmetry
breaking the super-heavy gauge bosons and their corresponding super-partners,
the gauginos, get a mass of order $M_G$ by ``absorbing" the would-be Goldstone
bosons $u^c_H,\,{\bar u}^c_H,\,e^c_H, \,{\bar e}^c_H$, and a linear
combination\footnote{$^{2}$}{The corresponding orthogonal combination of
$\nu^c_H$ and
${\bar\nu}^c_H$ remains ``uneaten", but still gets a mass of order $M_G$.} of
$\nu^c_H$ and ${\bar\nu}^c_H$. Moreover, also the fermionic components of $\bf
H$ and $\bf\bar{H}$, and in particular of  $\nu^c_H$ and ${\bar\nu}^c_H$,
forming massive states with the heavy gauginos, get a large mass at the GUT
scale.

The subsequent electroweak breaking down to $SU(3)_c\otimes U(1)_{em}$ is
induced spontaneously by the VEVs of two Higgs superfields ${\bf h(5,-2)}\sim
(D_3; h^o,h^-)$ and ${\bf{\bar h}({\bar 5},2)}\sim
({\bar D}_3; {\bar h}^+,{\bar h}^o)$, which form a 10-dimensional
representation of $SO(10)$; we shall set $<h>=<h^o>\equiv v_d\,$ and $\,<{\bar
h}>=<{\bar h}^o>\equiv v_u$, such that $(v_u^2+v_d^2)^{1/2}=v_{SM}= 246$ GeV.
One of the nice features of the model, as we have already mentioned above, is
the natural solution of the so-called doublet-triplet mass-splitting problem.
This is related to the fact that if the colour-triplet scalar fields $D_3$ and
${\bar D}_3$ have a mass of order $M_W$ (like their $SU(2)_L$-doublet
partners $(h^o,h^-)$ and $({\bar h}^+,{\bar h}^o)\,$), they would mediate a
too fast proton decay, in contradiction with the present experimental bounds.
In
order to solve this problem, also present in the standard GUTs, one needs to
find a mechanism for inducing a large mass of order $M_G$ to these
colour-triplets, while leaving unaffected the mass of the standard-type Higgs
doublets. A mechanism of this sort is naturally implemented in the flipped
model since $D_3$ and ${\bar D}_3$ form massive states at the GUT scale with
the
(uneaten) colour-triplets $d^c_H$ and ${\bar d}^c_H$ of the Higgs superfields
$\bf H$ and $\bf\bar{H}$.

The superfield content of the model is then completed by a set of $n_g+1\,$
$SU(5)$-singlets $\phi_\alpha$ ($n_g=3$ is the
number of generations) which may naturally arise in string theories, and
which, mixing with the RH neutrinos $\nu^c$,  are essential for providing the
desired seesaw mechanism.  The scalar component of one of
these singlets, denoted by $\phi_o$, develops a VEV at the weak scale. This,
apart from giving a mass to all $\phi$'s, is also necessary for producing the
correct mixing between the two electroweak Higgs doublets, insuring the
non-vanishing of both $v_u$ and $v_d$, and avoiding therefore the presence of
an unacceptable electroweak axion.

Finally, the model is completely specified by writing the most general
superpotential, which has to satisfy the discrete $Z_2$ symmetry ${\bf H}
\rightarrow - {\bf H}$ in order to prevent unplausible tree-level Majorana
masses (of order $M_G$) for the ordinary left-handed (LH) neutrinos:

$$\eqalign{{\cal W}_{(5)}  = &  \lambda_1^{ij} F_i F_j h +
\lambda_2^{ij} F_i {\bar f}_j {\bar h} +
\lambda_3^{ij} {\bar f}_i l_j^c h +
\lambda_4 H H h + \lambda_5 {\bar H} {\bar H} {\bar h} \cr & +
\lambda_6^{i \alpha} F_i {\bar H} \phi_{\alpha}  +
\lambda_7^{\alpha} h {\bar h} \phi_{\alpha} +
\lambda_8^{\alpha\beta\gamma} \phi_{\alpha} \phi_{\beta}
\phi_{\gamma}\,.\cr}\eqno(1)$$

\noindent Notice that this superpotential contains all possible terms which
satisfy the $SO(10)$ symmetry. After the electroweak symmetry breaking the
first three terms generate masses for all fermions, including Dirac masses for
the neutrinos. More precisely, the corresponding mass matrices may be written
as follows:

$$\eqalignno{m_d &= \lambda_1\, <h^o> \equiv \lambda_1\, v_d\,,&(2a)\cr
 m_u &= m_{\nu D} = \lambda_2\, <{\bar h}^o> \equiv \lambda_2\,v_u\,,&(2b)\cr
m_e &=  \lambda_3\, <h^o> \equiv \lambda_3 \,v_d\,,&(2c)\cr}$$

\noindent where the $\lambda_i$'s ($i$=1,2,3) are the $3\times 3$ matrices of
the Yukawa-type couplings, and $m_{\nu D}$ is the Dirac mass matrix for the
neutrinos. Notice the lack of any mass correlation among the different charged
fermion sectors, and in particular of the Standard-$SU(5)$ relations
$m_{d_i}(M_G) = m_{e_i}(M_G)$. The $\lambda_4-$ and the $\lambda_5-$terms give
rise to the necessary doublet-triplet mass-splitting discussed above,
producing mass terms of the type $\lambda_4 M_G (D_3 d^c_H)$ and
$\lambda_5 M_G ({\bar D}_3 {\bar d}^c_H)$. The $\lambda_6-$term produces the
mixing between the singlets\footnote{$^{3}$}{As in the previous
papers [3,4], we assume for simplicity that the fermionic partner of the
singlet
$\phi^o$ does not mix with the $\nu^c_i$.} $\phi_{\alpha}$ ($\alpha$=1,2,3) and
the RH neutrinos $\nu^c_i$. The $\lambda_7-$term gives rise to the correct
mixing between the two electroweak Higgs doublets, and the last
$\lambda_8-$term, which yields mass terms for the $\phi$'s of order $M_W$, is
necessary for preventing them from developing large VEVs. Furthermore, when
also the scalar partners of the RH neutrinos have a non-vanishing VEV,
$<\nu^c_i>$, the $\lambda_6-$term leads  to a mixing between the singlets
$\phi_{\alpha}$ and the heavy state ${\bar\nu}^c_H$:

$$\sum_{i=1}^3 \lambda_6^{i\alpha} <\nu^c_i>\,\,
(\phi_{\alpha}\,{\bar\nu}^c_H)\,,\eqno(3)$$

\noindent which will play a primary role in our discussion of the neutrino
mass matrix.

The study of the possible neutrino mass spectra which may arise in the context
of the flipped $SU(5)\otimes U(1)$ model has already been carried out by
several authors [3,4,6,7]. Here is a brief account of the results obtained more
recently\footnote{$^{4}$}{The Feynman diagrams which explain
the origin of neutrino masses in all cases have been shown in the Figs. of
ref.[4].}. At first , it was thought that the model could only lead to
uninteresting small neutrino masses of order $m_{\nu_i}\simeq (m_{u_i}/M_G)^2
M_W \leq 10^{-17}$ eV ($m_{u_i}$ being the mass of the up-quark of the $i$-th
generation), because of the so-called ``super-suppression" seesaw
mechanism, as a resulting of the mixing at the GUT mass scale of the RH
neutrinos with the singlets $\phi_{\alpha}$. Then, it was suggested [7] that
some of the neutrinos could get phenomenologically more interesting masses
given by  $m_{\nu_i}\simeq m_{u_i}^2/M_R \sim 10^{-4}$ eV to $100$ keV, as a
consequence of large Majorana masses for the RH neutrinos of order $M_R\simeq
10^8$ GeV, induced radiatively at the 2-loop level via the so-called Witten
diagram [8].  Nevertheless, as was remarked
in ref.[3], such a mechanism cannot be used in the more interesting (from the
point of view of the string) supersymmetric case, because of the
SUSY protection of the gauge hierarchy which prevents the radiative generation
of masses larger than the SUSY breaking scale [9], in the TeV range. On the
other hand, it was suggested [3] that even in this SUSY case a large RH
Majorana
mass scale of order $M_R\simeq M_G^2/M_S \simeq 10^{14}$ GeV (where
$M_S$=$10^{18}$ GeV is the string unification scale) might be induced by
introducing suitable non-renormalizable terms, like $(F {\bar H} {\bar H}
F)/M_S\,$, which are allowed but not completely specified by the string theory
[10]. In this case the neutrino masses are expected to be in the range
$10^{-10}$ to $0.1$ eV, still interesting for example for the solution of the
solar neutrino problem \`a la MSW [11]. Finally, in ref.[4] it was shown that
the neutrino mass spectrum could drammatically change in presence of large
vacuum expectation values for the RH sneutrinos. In particular, it was
suggested
that in this way one could get phenomenologically interesting masses without
the
need of the non-renormalizable terms. However, the model was studied only at a
single-generation level, and
considering only the limit cases where either the effects of such VEVs, or the
usual ({\it i.e.}, for all $<\nu^c_i> = 0$) ``super-suppression" seesaw
mechanism, were dominant. In the present paper, instead, we shall be able to
derive a more general expression for the three-generation light neutrino mass
matrix.  Interestingly this formula will show that, at least as far as the
matrix of the couplings $\lambda_6$ is non-singular, the part of the neutrino
mass matrix due to the presence of non-vanishing  $<\nu^c_i>$ turns out to be
independent on the particular structure (in the generation space) of the
Yukawa-type couplings entering in the superpotential, depending only on the
size of these VEVs and on the up-quark masses. But even more interesting is the
unexpected fact that the particular structure of the resulting mass matrix,
having a rank equal to 1, may lead to only one neutrino mass proportional to
the $<\nu^c_i>$, independently on the number of these VEVs which are non-zero
and on the other arbitrary parameters of the model. In other words, unless all
$<\nu^c_i>$ are zero, the model predicts that only one neutrino may have a
phenomenologically interesting mass, while the other two  are expected to be
much lighter, having eventually a tiny mass induced by the super-suppression
seesaw mechanism.

Our starting point for the calculation of the effective mass matrix for the
ordinary light neutrinos is the full mass matrix for the neutral fermions
which has already been introduced in ref.[4]. Its structure may easily
be obtained by analyzing the role played by the different terms of the
superpotential given in eq.(1). In particular, dropping the
piece relative to the massive ($\sim M_G$) state $\nu^c_H$, since it does not
affect the light neutrino sector, we may write the full mass matrix in the
basis
($\nu_i,\,\nu^c_j,\,\phi_{\alpha},\, {\bar\nu}^c_H$), with $i,j,\alpha$=
1,2,3, as:

$${\cal M}_{\nu} = \left(\matrix{0&U&0&0\cr
U^T&\cdot&G&\cdot\cr
0&G^T&X&V\cr 0&\cdot&V^T&M_G\cr}\right)\,,\eqno(4)$$

\noindent where $U$ is the up-quark mass matrix and $G=
\lambda_6\,<{\bar\nu}^c_H>\equiv \lambda_6\,M_G$ is the $3\times 3$ matrix
describing the mixing between the RH neutrinos and the singlets
$\phi_{\alpha}$, assumed to be non-singular. The ``dots" stand for entries not
larger than the weak scale which do not play an essential role in our
discussion. $X=\lambda_8\,<\phi_o>\simeq \lambda_8\,M_W$ is the $3\times 3$
mass matrix of the singlet fields $\phi_{\alpha}$. $M_G$ in the lower-right
corner of eq.(4) corresponds to the typical mass at the GUT scale for the
fermionic partner of the Higgs field ${\bar\nu}^c_H$ after the spontaneous
symmetry breaking. The
new interesting piece of ${\cal M}_{\nu}$ is the $3\times 1$ submatrix $V$
describing the terms given in eq.(3), which is due to the presence of
non-vanishing VEVs for the scalar partners of the RH neutrinos. We notice that
it arises from the same term in the superpotential as $G$, and  may be written
as:

$$V = (\lambda_6)^T \xi\,,$$

\noindent proportional to the transposed-conjugated of $\lambda_6$ and to the
$3\times 1$ vector $\,\xi = (<\nu^c_1>,\,
<\nu^c_2>,\, <\nu^c_3>)^T$. Now we may proceed to the evaluation of the
 ``effective" mass matrix for the three
ordinary light neutrinos. Following a usual procedure employed in the
seesaw-type of models, we may write:

$$m_{\nu}\simeq (U\,\,\,0\,\,\,0)
\left(\matrix{\cdot&G&\cdot\cr G^T&X&V\cr \cdot&V^T&M_G\cr}\right)^{-1}
\left(\matrix{U^T\cr 0\cr 0\cr}\right)\,\equiv U\,(R^{-1})\,U^T\,,\eqno(5)$$

\noindent where, approximately (up to terms of order $<\nu^c_i>/M_G$), the
effective $3\times 3$ RH Majorana mass matrix $R$ may be
expressed as:

$$R\simeq (G\,\,\,0)
\left(\matrix{X&V\cr V^T&M_G\cr}\right)^{-1}
\left(\matrix{G^T\cr 0}\right)\,\equiv G\,(\Omega^{-1})\,G^T\,.\eqno(6)$$

\noindent Similarly, the effective $3\times 3$ matrix $\Omega$ can be obtained
by means of the formula:

$$\Omega\simeq X - {1\over M_G}\, V\, V^T\,.\eqno(7)$$

\noindent Using  eqs.(5-7) we can then re-write our effective mass
matrix for the three light neutrinos as:

$$m_{\nu}\simeq U\, (R^{-1})\, U^T\,\simeq U\, G^{T -1}\,\Omega\,
G^{-1}\,U^T\,\simeq (U\,G^{T -1})\, (X - {1\over M_G}\, V \, V^T)\, (G^{-1}\,
U^T)\,.\eqno(8)$$

\noindent Therefore, using the definitions of $G$ and $V$ given above, we get
finally:

$$m_{\nu}\simeq {1\over M_G^2}\, U\,\left\{ \left(\lambda_6^{T
-1}\,X\,\lambda_6^{-1}\right)\,-\,{1\over M_G}\,
\left(\xi\,\xi^T\right)\right\}\,U^T\,.\eqno(9)$$

\noindent This formula, which is our main result, gives the approximate
expression of the neutrino mass matrix in the more general case. As we
mentioned
 above, a remarkable feature
 is the cancellation of the explicit dependence of the second
term on the unknown structure of the Yukawa-type couplings $\lambda_6$. This
means therefore that we may get predictions for the non-negligibly small
neutrino masses, due to the presence of $<\nu^c_i>$, only specifying the
pattern of the VEVs for the RH sneutrinos. Of course we
are aware that the procedure used above for its derivation may  be not quite
appropriate, especially for $<\nu^c_i>$ of order $M_G$, but we
have checked through a numerical analysis that it leads to the correct
behaviour
for the neutrino mass spectrum. Furthermore, at a simple single-generation
level, the two terms in eq.(9) lead, respectively, to the formulas
already obtained in refs.[3,4] in the limits  $<\nu^c>\to 0$ and
$<\nu^c>\,\,\gg
\sqrt{M_W M_G}\simeq 10^9$ GeV. In fact, the first term corresponds to the
``super-suppression" seesaw, which typically gives masses $m_{\nu_i}\simeq
({m_u^2}_i/M_G)^2\,M_W\sim 10^{-25}$ to $10^{-17}$ eV, negligibly too small
for any phenomenological application. The second term, on the other hand,
is due to the non-vanishing of some of the $<\nu^c>$, and may lead to
larger neutrino masses given by $m_{\nu_i}\simeq
({m_u^2}_i\,<\nu^c_i>)^2/M_G^3$. In particular, for $<\nu^c>\simeq M_G$, it
gives $m_{\nu}\simeq m_u^2/M_G$, recovering the result of the standard-GUT
seesaw scenarios, with the scale of the RH Majorana masses set at
$M_G$. Nevertheless, a closer view of the structure of eq.(9)
shows that, independently on the form assumed for the up-quark mass matrix,
only one neutrino may get a mass proportional to
$<\nu^c>$. The reason for this strong statement, which implies therefore
that in the flipped model at least two neutrinos are expected to have
extremely small masses, is due to the structure of
the matrix in the second term of eq.(9). In fact, since it is given by the
product of a $3\times 1$ vector ($U\,\xi$) times its transposed-conjugated, it
may only have a rank equal to one, corresponding to only one non-zero
eigen-mass given by:

$$m_{\nu_3}\,\simeq \,{1\over M_G^3}\,\,|
U\,\xi\,|^2\,.\eqno(10)$$

\noindent In this equation $|\,U\,\xi\,|^2$ is the modulo-squared of the
$3\times 1$ vector obtained by  the product of the up-quark mass matrix with
the vector $\xi$ of the VEVs $<\nu^c_i>$. The other two neutrinos, on the
other hand, will in general get a very tiny mass, either through the first term
of eq.(9), or still from the second term, but at a higher order of
approximation.(and,
In other
words we predict that, unless all VEVs $<\nu^c_i>$ are set to zero, the light
neutrino spectrum in the flipped model is made by two  very light states and
one heavier neutrino with a mass given by eq.(10). As an example, assuming
for simplicity a diagonal up-quark mass matrix, one would get:

$$m_{\nu_3}\,\simeq {1\over M_G^3} \left\{\left(m_u\,<\nu^c_1>\right)^2 +
\left(m_c\,<\nu^c_2>\right)^2 + \left(m_t\,<\nu^c_3>\right)^2\,\right\}.$$

\noindent In particular, for $<\nu^c_3>\simeq M_G$, the third (say, $\tau$-)
neutrino mass will be given by $m_{\nu_3}\simeq m_t^2/M_G\simeq 10^{-3}$ eV, in
the correct range for solving the solar neutrino problem via the MSW
oscillations $\nu_e\rightarrow \nu_3$. We wish to stress here, once more, that
this result holds in the present model under quite general assumptions, namely
the non-singularity of the Yukawa-type matrix $\lambda_6$ and the presence of
VEVs for the RH sneutrinos at the GUT mass scale.

In conclusion, what we have done in the present paper, is a detailed analysis
of the light neutrino mass spectrum which may be obtained in the context of the
supersymmetric flipped $SU(5)\otimes U(1)$ model, obtaining a general
(approximate) formula for the corresponding three-generation mass matrix. The
particular structure of such expression has also suggested that the presence of
large ($\simeq M_G$) VEVs for the RH sneutrinos may at most lead to only one
neutrino mass in the range of phenomenological interest for astrophysics and
cosmology. The other two neutrinos, instead, are expected to have a too small
mass.

\vskip 2 cm
%\vfill
%\eject

\noindent {\bf References}

\item{[1]} I. Antoniadis, J. Ellis, J.S. Hagelin and D.V. Nanopoulos,
 Phys. Lett. {\bf 194B} (1987) 231;  {\bf 205B} (1988) 459;  {\bf 208B} (1988)
209; {\bf 231B} (1989) 65.

\item{[2]}  I. Antoniadis, G.K. Leontaris and J. Rizos, Phys.
Lett. {\bf 245B} (1990) 161.

\item{   } G.K. Leontaris, J. Rizos and K. Tamvakis,  Phys.
Lett. {\bf 243B} (1990) 220; {\bf 251B} (1990) 83;

\item{   } I. Antoniadis, J. Rizos and K. Tamvakis, Phys. Lett. {\bf 278B}
(1992) 257; {\bf 279B} (1992) 281;

\item{   } J.L. Lopez and D.V. Nanopoulos, Nucl. Phys. {\bf B338} (1990) 73;
Phys. Lett. {\bf 251B} (1990) 73.

\item{   } D. Bailin and A. Love, Phys. Lett. {\bf 280B} (1992) 26.

\item{[3]} E. Papageorgiu and S. Ranfone, Phys. Lett. {\bf 282B} (1992) 89.

\item{[4]} S. Ranfone and E. Papageorgiu, Phys. Lett. {\bf 295B} (1992) 79.

\item{[5]} S. Ranfone, Phys. Lett. {\bf 286B} (1992) 293.

\item{[6]} Alon E. Faraggi,  Phys. Lett. {\bf 245B} (1990) 435;

\item{   } I. Antoniadis, J. Rizos and K. Tamvakis, Phys. Lett. {\bf 279B}
(1992) 281;

\item{   } J. Ellis, J. Lopez and D.V. Nanopoulos,  Phys. Lett. {\bf 292B}
(1992) 189;

\item{   } J. Ellis, D.V. Nanopoulos and Keith A. Olive,  Phys. Lett. {\bf
300B} (1993) 121.

\item{   } G.K. Leontaris and J.D. Vergados, Phys. Lett. {\bf
305B} (1993) 242.

\item{[7]} G.K. Leontaris and J.D. Vergados, Phys. Lett. {\bf
258B} (1991) 111.

\item{[8]} E. Witten, Phys. Lett. {\bf 91B} (1980) 81.

\item{[9]} L. Ibanez, Phys. Lett.,{\bf 117B} (1982) 403;  Nucl.
Phys.,{\bf B218} (1983) 514.

\item{[10]} S. Kalara, J.L. Lopez and D.V. Nanopoulos, Phys.
Lett., {\bf 245B} (1990) 421;

\item{   } J.L. Lopez and D.V. Nanopoulos, Phys.
Lett., {\bf 251B} (1990) 73;

\item{   } J. Rizos and K. Tamvakis, Phys. Lett.,{\bf 262B} (1991) 227.

\item{[11]} S. Mikheyev and A. Smirnov,  Sov. J. Nucl.
Phys., {\bf 42} (1986) 1441;

\item{   } L. Wolfenstein, Phys. Rev., {\bf D17} (1978) 2369;
{\bf D20}  (1979) 2634.

\item{[12]} For a review on seesaw models for neutrinos, see, {\it e.g.},
S. Ranfone, RAL preprint (June 1992), RAL-92-039, {\sl
(unpublished)}.

\vfill\eject
\bye